\documentclass[amsmath,amssymb,aip,apl,showpacs,reprint,superscriptaddress]{revtex4-1}

\usepackage{graphicx}
\usepackage{dcolumn}
\usepackage{bm}
\usepackage{siunitx}
\usepackage[colorlinks=true, linkcolor=blue, citecolor=blue, urlcolor=blue]{hyperref}

\def\emph#1{\textcolor{red}{#1}}
\def\emph#1{\textcolor{black}{#1}}
\begin{document}


\title{Enhancement of intrinsic magnetic damping in defect-free epitaxial Fe$_3$O$_4$ thin films}


\author{Xianyang Lu}
\affiliation{York-Nanjing Joint Center (YNJC) for Spintronics and Nano-engineering, School of Electronic Science and Engineering, Nanjing University, Nanjing, 210093, China}
\affiliation{Department of Physics, University of York, York, YO10 5DD, UK}

\author{Lewis J. Atkinson}
\affiliation{Department of Physics, University of York, York, YO10 5DD, UK}

\author{Balati Kuerbanjiang}
\affiliation{Department of Physics, University of York, York, YO10 5DD, UK}

\author{Bo Liu}
\affiliation{York-Nanjing Joint Center (YNJC) for Spintronics and Nano-engineering, School of Electronic Science and Engineering, Nanjing University, Nanjing, 210093, China}

\author{Guanqi Li}
\affiliation{York-Nanjing Joint Center (YNJC) for Spintronics and Nano-engineering, School of Electronic Science and Engineering, Nanjing University, Nanjing, 210093, China}
\affiliation{Department of Physics, University of York, York, YO10 5DD, UK}

\author{Yichuan Wang}
\affiliation{York-Nanjing Joint Center (YNJC) for Spintronics and Nano-engineering, School of Electronic Science and Engineering, Nanjing University, Nanjing, 210093, China}
\affiliation{Department of Physics, University of York, York, YO10 5DD, UK}

\author{Junlin Wang}
\affiliation{York-Nanjing Joint Center (YNJC) for Spintronics and Nano-engineering, School of Electronic Science and Engineering, Nanjing University, Nanjing, 210093, China}
\affiliation{Spintronics and Nanodevice Laboratory, Department of Electronic Engineering, University of York, York YO10 5DD, UK}

\author{Xuezhong Ruan}
\affiliation{York-Nanjing Joint Center (YNJC) for Spintronics and Nano-engineering, School of Electronic Science and Engineering, Nanjing University, Nanjing, 210093, China}

\author{Jing Wu}
\email[Email:~]{jing.wu@york.ac.uk}
\affiliation{York-Nanjing Joint Center (YNJC) for Spintronics and Nano-engineering, School of Electronic Science and Engineering, Nanjing University, Nanjing, 210093, China}
\affiliation{Department of Physics, University of York, York, YO10 5DD, UK}

\author{Richard F. L. Evans}
\affiliation{Department of Physics, University of York, York, YO10 5DD, UK}

\author{Vlado K. Lazarov}
\affiliation{Department of Physics, University of York, York, YO10 5DD, UK}

\author{Roy W. Chantrell}
\affiliation{Department of Physics, University of York, York, YO10 5DD, UK}

\author{Yongbing Xu}
\email[Email:~]{yongbing.xu@york.ac.uk}
\affiliation{York-Nanjing Joint Center (YNJC) for Spintronics and Nano-engineering, School of Electronic Science and Engineering, Nanjing University, Nanjing, 210093, China}
\affiliation{Spintronics and Nanodevice Laboratory, Department of Electronic Engineering, University of York, York YO10 5DD, UK}

\date{\today}

\begin{abstract}
We have investigated the magnetic damping of precessional spin dynamics in defect-controlled epitaxial grown Fe$_3$O$_4$(111)/Yttria-stabilized Zirconia (YSZ) nanoscale films by all-optical pump-probe measurements. The intrinsic damping constant of the defect-free Fe$_3$O$_4$ film is found to be strikingly larger than that of the as-grown Fe$_3$O$_4$ film with structural defects. We demonstrate that the population of the first-order perpendicular standing spin wave (PSSW) mode, which is exclusively observed in the defect-free film under sufficiently high external magnetic fields, leads to the enhancement of the magnetic damping of the uniform precession (Kittel) mode. We propose a physical picture in which the PSSW mode acts as an additional channel for the extra energy dissipation of the Kittel mode. The energy transfer from Kittel mode to PSSW mode increases as in-plane magnetization precession becomes more uniform, resulting in the unique intrinsic magnetic damping enhancement in the defect-free Fe$_3$O$_4$ film.
\end{abstract}


\maketitle
The photo-induced precessional spin dynamics in various magnetic materials has attracted significant attention since the observation of the uniform magnetic precession (Kittel mode) and the corresponding first-order perpendicular standing spin wave (PSSW mode) in Ni films by the all-optical pump-probe technique.\cite{1,3} After excitation by a femtosecond laser pulse, besides the uniform Kittel mode, different spin wave modes can be stimulated including first-order PSSW and Damon-Eshbach dipolar surface spin waves (DE modes).\cite{4} At the same time, all-optical pump-probe measurements allow determination of the magnetic Gilbert damping $\alpha$,\cite{6,7} which is a key parameter for magnetic data recording and the next-generation spintronic memory devices such as Magnetoresistive Random Access Memory (MRAM)~\cite{8,41}. Therefore, understanding and controlling the magnetic damping is of crucial importance. Among many factors affecting the magnetic damping, structural defects are crucial because they are generally inevitable when preparing films or devices. It was proposed theoretically that defects scatter the Kittel mode into short wavelength spin waves via two magnon scattering, producing an extrinsic contribution to magnetic damping.\cite{42} This extrinsic mechanism was verified by the fact that in thin NiFe films the FMR linewidth increases with decreasing thickness.\cite{43} Also, the magnetic damping was found to significantly increase due to interfacial defects in ultrathin Fe/Cr layers.\cite{44} {\color{blue}It is reported that the antisite disorder can increases the intrinsic Gilbert damping in \textit{L}1$_0$ FePt films in which the spinflip scattering plays dominant role.\cite{ma2015}} Thus, it can be expected that the magnetic damping should become smaller with reduced structural defects. {\color{blue}Although the damping coefficient is inversely proportional to the Co layer thickness in Co/Pd multilayers,\cite{pal2011}} a thickness-dependent study of magnetic damping in Fe$_3$O$_4/$MgO films has shown a strong increase of effective damping from $0.037$ up to $0.2$ with increasing film thickness from $\SI{5}{nm}$ up to $\SI{100}{nm}$,\cite{28} and the mechanisms of this observation is unknown. As a half metal, Fe$_3$O$_4$ is one of the promising materials for spintronics due to nearly 100\% spin polarization and highly efficient spin injection and transport. Fe$_3$O$_4$ contains defects including antiphase boundaries (APBs) and twin defects.\cite{wong2010,hassan2009} APBs are structural defects formed during the film growth originating from the mismatch between Fe$_3$O$_4$ films and substrates.\cite{15} The low formation energy of APBs means that they are commonly observed in thin film Fe$_3$O$_4$ samples.\cite{16,17} In addition to APBs, it was demonstrated that as-grown films can also have significant numbers of twin defects.\cite{18} The influence of the defects in Fe$_3$O$_4$ thin films has not been investigated and very few studies on the magnetization dynamics and magnetic damping of Fe$_3$O$_4$ thin films have been reported.\cite{26,27}

To explore the effects of the structure on magnetic damping, the defect-controlled epitaxial Fe$_3$O$_4$(111) nanoscale films on Yttria-stabilized Zirconia (YSZ) substrate were prepared by PLD growth and annealing. Using time-resolved magneto-optical Kerr effect (TR-MOKE), we find that the PSSW mode only exists in the annealed defect-free Fe$_3$O$_4$ film{\color{blue}, and numerical simulations support this finding.} More importantly, in contrast to the general belief, the magnetic damping of the defect-free Fe$_3$O$_4$ film is found to be significantly larger than that of the as-grown film with defects. We find that the non-uniform PSSW mode is an extra energy dissipation channel in addition to the uniform Kittel mode in the defect-free Fe$_3$O$_4$ film. The PSSW mode is only populated when the in-plane magnetization precession is uniform under high magnetic field. Our work demonstrates that the PSSW mode draws energy from the Kittel mode and then leads to the enhancement of the magnetic damping of the uniform magnetic precession.

\begin{figure}
	\includegraphics[width=3.375in]{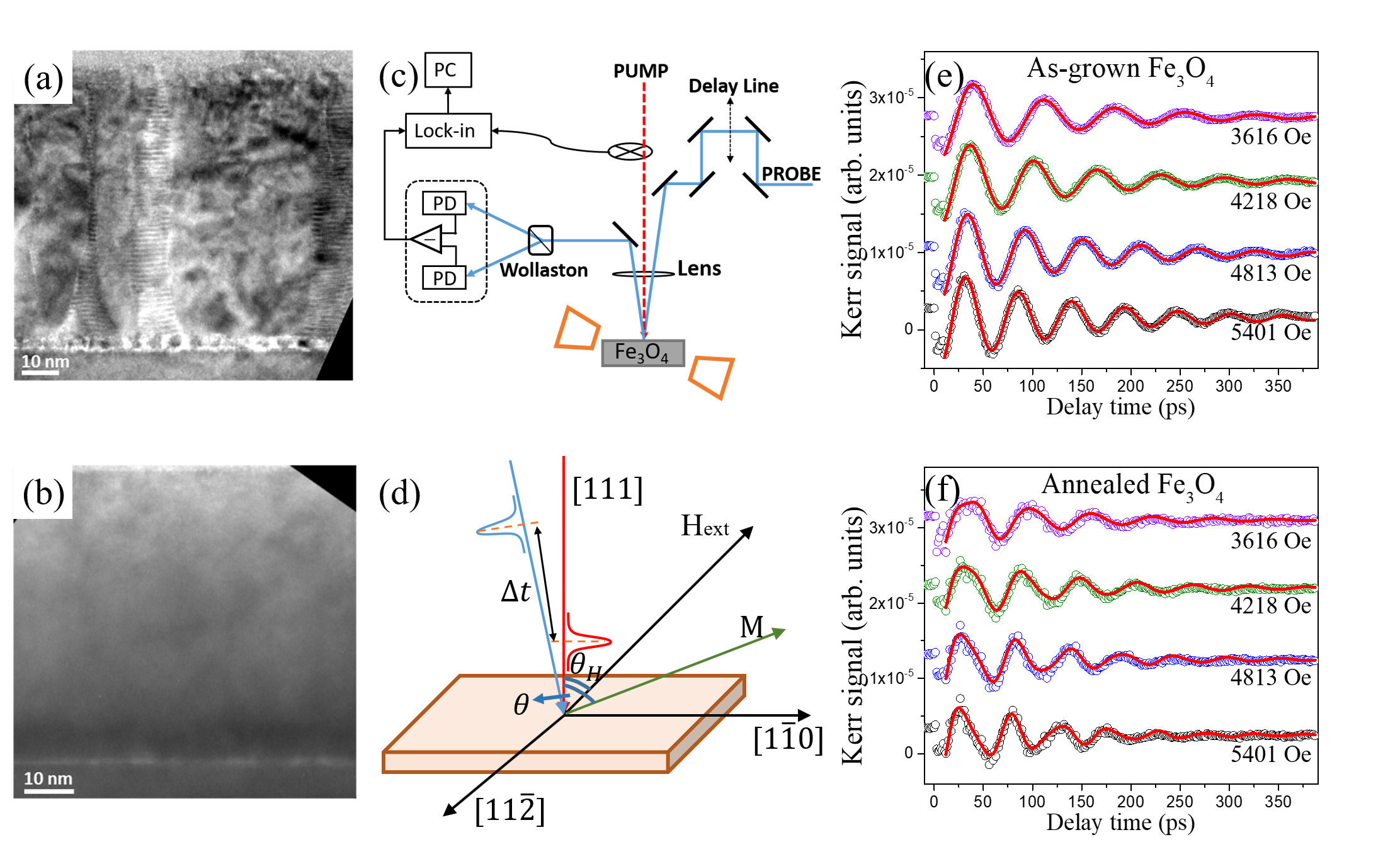}
	\caption{\label{fig1}(a)(b) TEM images of (a) the as-grown Fe$_3$O$_4$/YSZ$\left(111\right)$ interface and (b) the annealed Fe$_3$O$_4$/YSZ$\left(111\right)$ interface. (c)(d) Schematic pump-probe TR-MOKE setup and sample geometry. (e)(f) Measured transient Kerr signal of as-grown Fe$_3$O$_4$ (e) and annealed Fe$_3$O$_4$ (f). Kittel mode are observed in both films while the PSSW mode is exclusively observed in annealed Fe$_3$O$_4$. The red solid line in (e) and (f) show the best fitting using single and double damped sinusoidal formulas, respectively. Applied field range is from $\SI{3616}{Oe}$ to $\SI{5404}{Oe}$.}
\end{figure}

To compare the defect conditions between the two as-grown and annealed films, transmission electron microscopy (TEM) observations have been performed. We find that the annealed Fe$_3$O$_4$ film contains a lower density of defects, including APBs and twin defects, compared to the as-grown film. The cross-sectional TEM images of as-grown and annealed Fe$_3$O$_4$ films are shown in Fig.~\ref{fig1}(a) and Fig.~\ref{fig1}(b), respectively(details in supplemental materials). {\color{blue}The thickness $L$ of as-grown film and annealed films are about $\SI{80}{nm}$  and $\SI{50}{nm}$, respectively.} 

The schematic pump-probe TR-MOKE setup and the sample geometry in the external field are illustrated in Fig.~\ref{fig1}(c) and (d). TR-MOKE results for both as-grown and annealed Fe$_3$O$_4$ samples are presented in Fig.~\ref{fig1}(e) and (f) under a pump beam fluence of $\SI{5.09}{\milli\joule\per\square\centi\metre}$. The photo-induced magnetic precession is described using a phenomenological formula:$\Delta\theta_{K}\propto\sum_{i=1}^{2}A_{i}\exp{\left(-t/\tau_{i}\right)\sin{\left(2\pi f_{i}t+\varphi_{i}\right)}+B\left(t\right)}$ where $\tau_{i}$ and $f_{i}$ are the relaxation time and the precession frequency, respectively. $\varphi_{i}$ is the initial phase of magnetization precession and $B\left(t\right)$ represents the background accounting for the slow recovery of the magnetization, which is very weak in our measurements due to the low pump beam fluence. A beating can be clearly seen from the Kerr rotation oscillations of the annealed film in Fig.~\ref{fig1}(f)(open circles). Therefore, a sum of two damped sinusoidal functions is used to give the best fit of the data as shown in solid lines in Fig.~\ref{fig1}(f). In the case of the as-grown sample, only a single damped sinusoidal function ($A_{2}=0$) is needed in order to give the best fit (solid lines) to the results (open circles) as shown in Fig.~\ref{fig1}(e). The best fit curves demonstrate the fitting function a good representation of the experimental results. The first dominating mode ($i=1$) observed in both films is the magnetic uniform precession mode obeying the Landau-Lifshitz-Gilbert (LLG) equation while the second mode ($i=2$), which only presents in the annealed film, is the first-order PSSW. We note that the possibility of a DE mode is excluded as discussed in supplemental materials.


\begin{figure}
	\includegraphics[width=3.1in]{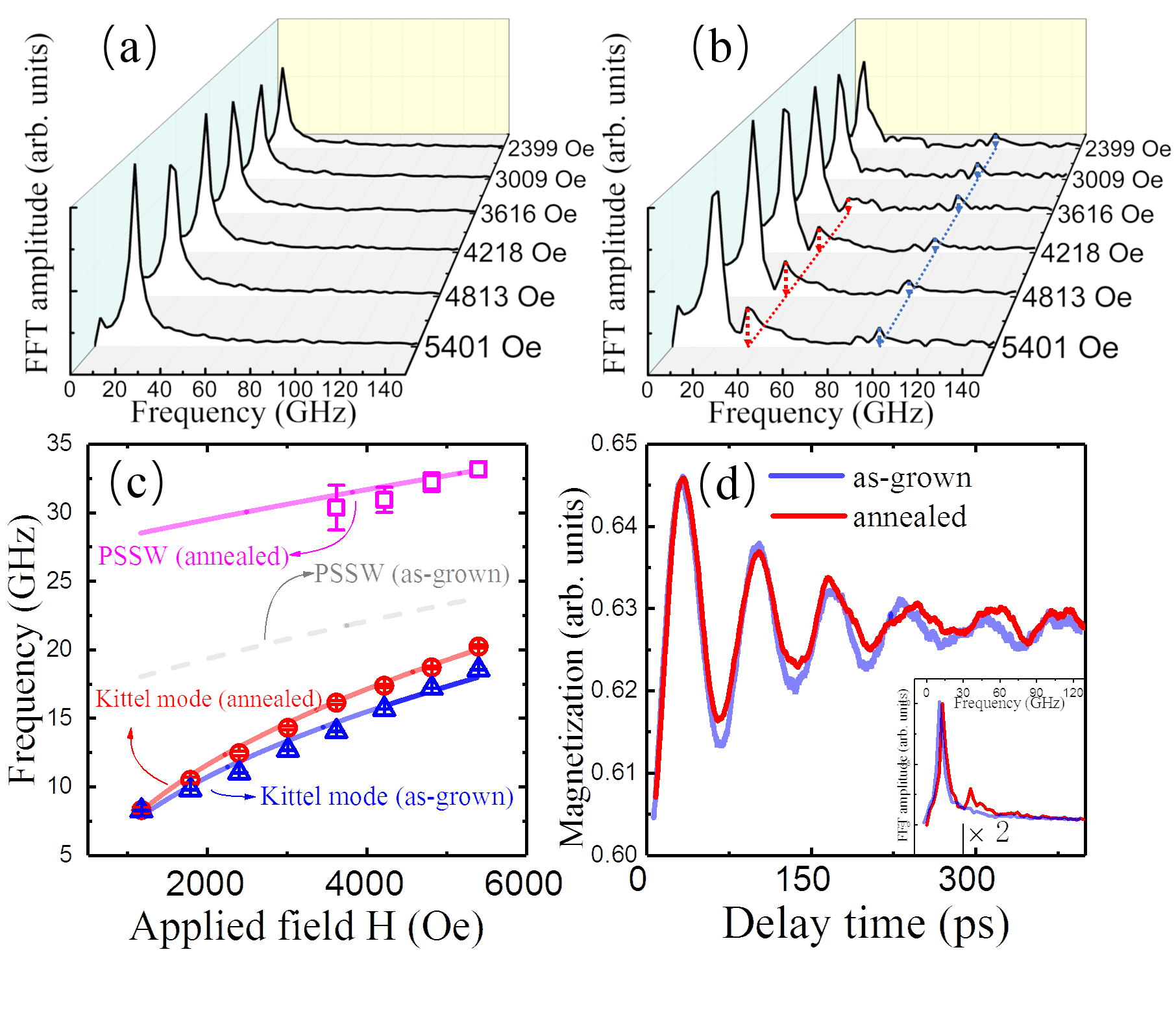}
	\caption{\label{fig2}Fourier spectrum of as-grown (a) and annealed (b) Fe$_3$O$_4$. Kittel mode is recognized as the main peak. In annealed film, the second mode (red dashed line) originate from the PSSW. The field independent high-frequency third mode is CAP mode (blue dashed line) from the Fe$_3$O$_4/$YSZ interface. (c) $f_1$ of Kittel mode of both as-grown film (blue triangles) and annealed film (red circles). $f_2$ of the PSSW mode observed in the annealed film is also plotted (pink squares). The solid lines represent the fitting curves. Dashed grey line represents the estimated PSSW mode in as-grown film which is not observed. (d) Numerical simulation of magnetization dynamics in as-grown (blue) and anneled (red) Fe$_3$O$_4$ thin film. Inset is the corresponding Fourier spectrums.}
	
\end{figure}

Fourier analysis is used to explore the observed multiple-mode oscillations which are presented in Fig.~\ref{fig2}. Fig.~\ref{fig2}(a) presents the Fourier spectrum for the as-grown film showing a single precession mode, the Kittel mode. Fig.~\ref{fig2}(b) presents the Fourier spectrum of the annealed film. In addition to the main peak denoting the Kittel mode (1st mode), two extra high-frequency modes are observed. The second mode (red dashed line), which only becomes profound at high external fields, is the PSSW mode. The amplitude of the Fourier spectrum of the PSSW mode increases as the external magnetic field increases, indicating that the excitation of PSSW modes gets stronger as the sample magnetization becomes more uniform. The frequency of the Kittel mode in the as-grown film is plotted as a function of the external field strength in blue triangles in Fig.~\ref{fig2}(c). The frequencies of the Kittel mode and PSSW mode in the annealed film are plotted vs the field strength in red circles and pink squares in Fig.~\ref{fig2}(c), respectively. The third mode (blue dashed line in Fig.~\ref{fig2}(b)) with a frequency of $~\SI{95.8}{GHz}$ shows no magnetic-field-dependence, therefore is not of a magnetic origin. This third mode is a coherent acoustic phonon (CAP) mode from the interface between the Fe$_3$O$_4$ film and substrat. The non-appearance of the CAP mode in the as-grown film may be due to phonon diffraction caused by defects. The Kittel and the PSSW frequencies in Fig.~\ref{fig2}(c) are fitted using the Eq. (S8) and Eq. (S10) derived from the LLG equation. The fitting processes are detailed in the supplemental material. The fits are in agreement with the values (data points) extracted from damped sinusoidal fitting and plotted in solid lines in Fig.~\ref{fig2}(c). The best fitted values of the perpendicular anisotropy constant $K_{u}$ and cubic anisotropy constant $K_{c}$ are obtained for both samples and are comparable with other studies on Fe$_3$O$_4$ thin films.\cite{27,31} The extracted value of the exchange stiffness constant $A_{ex}$ from the PSSW mode fitting is $\SI[separate-uncertainty = true]{1.47(15)}{\micro erg\per\centi\meter}$ which is reasonably close to the reported value of $A_{ex}=\SI{1.19}{  \micro erg\per\centi\meter}$ estimated for exchange coupling constant of Fe$_3$O$_4$.\cite{32} If we substitute $K_{u}$, $K_{c}$, $L$ and $M_{s}$ of as-grown film as well as the extracted $A_{ex}$ from annealed film into Equation (S10), the field dependence of the frequency of the PSSW mode in the as-grown film is estimated and plotted as the dashed line in Fig.~\ref{fig2}(c). The difference in the PSSW mode between two samples comes from the different film thickness between them. However, this PSSW mode is not observed, neither in the TR-MOKE results (Fig.~\ref{fig1}(e)) nor the corresponding Fourier spectrum (Fig.~\ref{fig2}(a)). Therefore, the PSSW mode is only observed in the annealed Fe$_3$O$_4$ film and is not present in the as-grown film due to magnon scattering induced by a high density of defects in this sample. The fact that the PSSW mode is not observed in the low field region can be attributed to magnon scattering due to the magnetic inhomogeneities. In support of the experimental data a model based on the Landau-Lifshitz-Bloch (LLB)\cite{33,34,35,36} equation has been used to provide further insight into the cause of the additional peaks seen in the Fourier spectrum. 

\begin{figure}
	\includegraphics[width=3.375in]{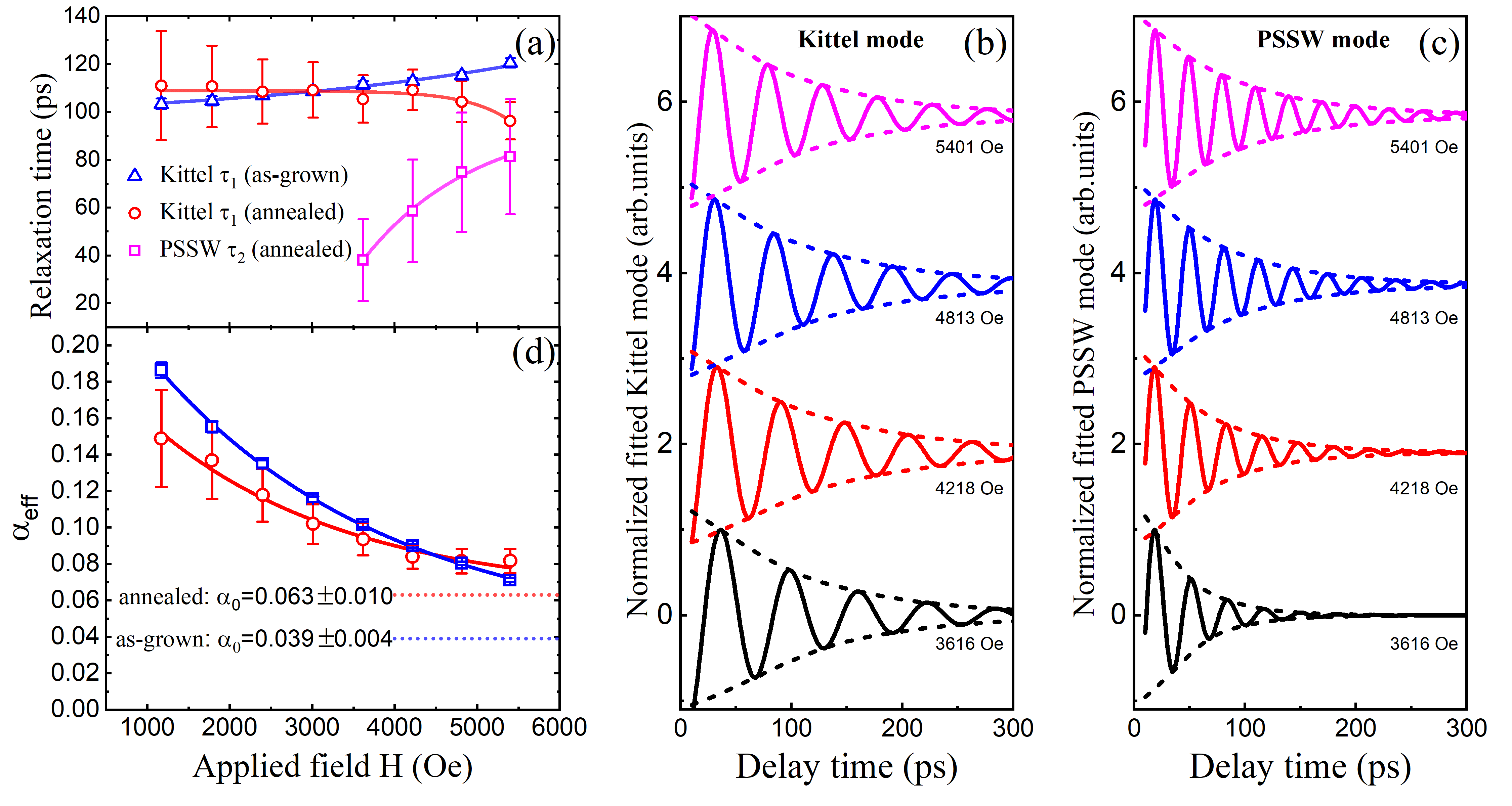}
	\caption{\label{fig3} Extracted $\tau_{1}$ of Kittel mode from as-grown Fe$_3$O$_4$ film (blue triangles) and annealed Fe$_3$O$_4$ film (red circles), $\tau_2$ of PSSW mode from the annealed film (pink squares). Solid lines are guide to eyes. (b)(c) Normalized time-domain fitting curves of PSSW mode (c) and Kittel mode (d), dashed lines are the envelop curves to monitor the amplitude decay. (d) $\alpha_{\mathrm{eff}}$ of as-grown (blue) and annealed (red) Fe$_3$O$_4$ films. From the single exponential decay fitting (solid lines), the intrinsic damping $\alpha_0$ of both films are estimated.}
\end{figure}

The relaxation time $\tau_{1}$ of the Kittel mode for both as-grown and annealed samples is obtained from the best fit of Equation (1) as a function of external field strengths and presented as blue triangles and red circles in Fig.~\ref{fig3}(a), respectively. The relaxation time $\tau_{2}$ of the second mode, the PSSW mode, observed in the annealed sample is also obtained and presented as pink squares in Fig.~\ref{fig3}(a). The PSSW mode emerges in the annealed film only when the applied field is sufficiently large. In Fig.~\ref{fig3}(a), the relaxation time of the Kittel mode, $\tau_{1}$ , increases with the field strengths in the as-grown sample. This is as expected since the relaxation time {\color{blue}is negatively correlated to} the energy dissipation rate of each spin wave modes, and the energy dissipation rate decreases when the magnetization becomes more uniform. In the annealed film, however, the relaxation time of the Kittel mode decreases with the field strengths. This is in contrast with the observation in the as-grown film. There are less lattice defects in annealed film, which means that the contribution of magnon scattering on the energy dissipation rate is not as strong as that in the as-grown film. However, one still expects that as the external field increases, the magnetization precession becomes more uniform and leads its energy dissipation rate decreasing rather than increasing. The fact that the precessional energy dissipates quicker as the increased field strength indicates that there must be an additional channel. This additional channel is responsible for the extra energy dissipation from the Kittel mode, and this channel draws more energy from the Kittel mode as the external field strength increases. On the other hand, PSSW modes are observed only in the annealed sample. The relaxation time, $\tau_{2}$, of the PSSW mode increases dramatically with the external field strength while $\tau_{1}$ of the Kittel mode decreases. This suggests that the PSSW mode acts as the additional channel responsible for the extra energy dissipation of the Kittel mode. In this case, the energy is transferred from the Kittel mode to the PSSW mode during magnetization precession in the annealed sample. This energy transfer between two modes increases as the in-plane magnetization precession becomes more uniform. {\color{blue}It should be noted that the relaxation time of Kittel model decreases with a larger field for the annealed sample may also result from that the effective damping reaches the intrinsic damping value as shown in the following discussion. Even though, the PSSW mode acting as an additional energy dissipation channel is clarified.}

The effective damping constants $\alpha_{\mathrm{eff}}$ are derived from the Kittel mode using~\cite{11,38} $\alpha_{\mathrm{eff}}=1/2\pi f_{1}\tau_{1}$, where $f_{1}$ and $\tau_{1}$ are the frequency and relaxation time of Kittel mode, respectively. The variations of the damping with the external field strength for the as-grown and annealed films are shown in Fig.~\ref{fig3}(d). The effective damping constant $\alpha_{\mathrm{eff}}$ in both films decreases significantly with increasing applied field. This field dependence of the effective damping is commonly observed and explained. The effective damping constant consists of intrinsic and extrinsic components and the extrinsic damping mainly comes from magnetic inhomogeneity or two magnon scattering. In the low magnetic field region, the effective field is dominated by the spatially fluctuating anisotropy field which leads to increased damping, while in the high field region the effect of anisotropy field becomes weak since the external field dominates and hence the effect of the magnetic inhomogeneity decreases along with the damping constant.\cite{29,39,40} In our measurement, the effective damping constant of the as-grown film is larger than that of the annealed Fe$_3$O$_4$ film at small external field strengths. This is consistent with the larger damping associated with lattice defects introduced by APBs or other defects present in the as-grown film. However, an intriguing finding is that as the external field strength becomes larger, $\alpha_{\mathrm{eff}}$ of the as-grown film becomes smaller than $\alpha_{\mathrm{eff}}$ of the annealed film. In light of the fact that the measured effective damping parameter usually decreases dramatically with the increasing of the applied field and eventually reaches a constant value,\cite{39,29,liu2018, qiao2013} a phenomenological fitting using single exponential decay is applied to describe the field dependence of $\alpha_{\mathrm{eff}}$ as shown in $\alpha_{\mathrm{eff}}=\alpha_{0}+\alpha_{\mathrm{ext}}\exp{(-\beta H)}$,
where $\alpha_{0}$ denotes the intrinsic damping. The second term $\alpha_{\mathrm{ext}}\exp{(-\beta H)}$ represents the extrinsic damping term which is dependent to the applied field. The best fittings as shown in Fig.~\ref{fig3}(d) give the value $\alpha_{0}=0.039\pm 0.004$ for the as-grown and $\alpha_{0}=0.063\pm 0.010$ for the annealed film. The intrinsic damping constant $\alpha_{0}$ of the defect-free annealed Fe$_3$O$_4$ film is thus 62\% higher than that of the as-grown film with high density of defects.

\begin{figure}
	\includegraphics[width=2.1in]{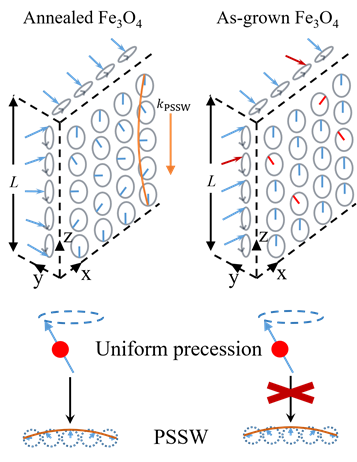}
	\caption{\label{fig4} Left panel: Without defect-induced magnon scattering, the PSSW is observed in annealed Fe$_3$O$_4$ film along with coherent Kittel magnetic precession. The Gilbert damping is enhanced due to the additional energy dissipation channel from Kittel mode to PSSW mode. Right panel: only the Kittel mode is observed because of the existence of magnetic inhomogeneity (red symbols). The energy dissipation channel from Kittel mode to PSSW mode no longer exists leading to a smaller damping compared to the defect-free film.} 
\end{figure}

As discussed, the energy is transferred from the Kittel mode to the PSSW mode at large external fields. To visualise this energy transfer, the time sequences of the Kittel and PSSW modes under the highest external field strengths are simulated using the extracted parameters (relaxation time, frequencies and initial phases), and plotted in Fig.~\ref{fig3}(c) and (d), respectively. Comparing the two sets of time sequences from bottom to top as the field increases from $\SI{3616}{Oe}$ to $\SI{5401}{Oe}$ in Fig.~\ref{fig3}(c) and (d), the decay of the PSSW mode slows down while the decay of the Kittel mode gradually speeds up. Without the PSSW mode, the decay of the Kittel mode is expected to decrease with the increasing of the external field strength, as in the case of the as-grown film. The existence of the PSSW mode has reversed this trend by providing an extra energy dissipation channel, which contributed to the intrinsic damping in the annealed Fe$_3$O$_4$ film. We therefore propose that this unexpected enhancement of magnetic damping in the annealed film is related to the emerging of the first order PSSW mode, which draws the energy from the Kittel mode and speeds its relaxation process. {\color{blue} This energy transfer process may be attributed to two potential mechanisms: nonlinear spin-waves transition\cite{5} and two-magnon scattering\cite{zakeri2007}.} The physical picture to explain the difference in magnetic damping of these two Fe$_3$O$_4$ films is illustrated in Fig.~\ref{fig4}. In the as-grown film, only the Kittel mode exists and PSSW modes isn’t populated even in high field region due to the magnetization inhomogeneity caused by defects. The magnetic damping follows the usual trend and approaches to its intrinsic value as the extrinsic contribution reduces in the high field region. In the annealed film with no structural defects, a first order PSSW mode is populated at the high field region. The amplitude of this PSSW mode increases and its relaxation time decreases with the external field strength, which indicates the energy drawn into the PSSW mode increases.  This suggests that the uniformity of in-plane magnetization is the key to populate higher energy magnon, PSSW modes,\cite{25} across the film thickness. The more uniform the magnetization is, the more energy the PSSW mode draws from the Kittel mode, which leads to a larger intrinsic damping of the Kittel mode in the defect-free films. 

In summary, the dynamic damping properties of the epitaxial grown Fe$_3$O$_4$ thin films with controlled atomic scale structures have been studied by an all-optical TR-MOKE technique. In addition to the dominate uniform magnetization precession Kittel mode, the first-order PSSW mode is observed exclusively in the annealed film with low density of defects when the applied field is sufficiently strong. The amplitude of the PSSW mode increases and its relaxation time decreases as the external magnetic field increases. Furthermore, the extracted intrinsic magnetic damping constant of the defect-free film is much larger than that of the film with defects. This enhancement is attributed to the PSSW mode, which provides the additional channel for the extra energy dissipation of the Kittel mode. During the magnetization precession of the defect-free Fe$_3$O$_4$ film, energy is transferred from the Kittel mode to the PSSW mode and this energy transfer between two modes increase as in-plane magnetization precession become more uniform resulting in the enhanced intrinsic magnetic damping. Our work demonstrates that the defect-free film structure is essential for the population of the PSSW modes and at the same time the existence of these PSSW modes provides an additional channel for the energy dissipation of the Kittel mode, which leads to the enhanced intrinsic magnetic damping in the epitaxial defect-free Fe$_3$O$_4$ thin films.  This result offers new insights into engineering the magnetic damping for potential spintronics applications employing Fe$_3$O$_4$ as a function material.

\bigskip
See supplemental material for details of sample preparation, TR-MOKE setup, TEM images, data fitting process, exchlusion of DE mode, numeriacal simulations, hysterisis loops and CAP mode.
\bigskip

\begin{acknowledgments}
	This work was supported by the National Basic Research Program of China (No. 2014CB921101), National Natural Science Foundation of China (No. 61274102, No. 61427812 and No. 11574137), The National Key Research and Development Program of China (No. 2016YFA0300803), Jiangsu NSF (BK20140054), Jiangsu Shuangchuan Programme. RFLE acknowledges the financial support of the Engineering and Physical Sciences Research Council (Grant No. EPSRC EP/P022006/1).
\end{acknowledgments}

\bibliography{Fe3O4}

\end{document}